\pdfoutput=1

\documentclass[11pt]{article}

\usepackage{acl}

\usepackage{times}
\usepackage{latexsym}

\usepackage[T1]{fontenc}

\usepackage[utf8]{inputenc}

\usepackage{microtype}

\usepackage{comment}
\usepackage{breqn}
\usepackage{caption}
\usepackage{graphicx}
\usepackage{booktabs}
\usepackage{xcolor}
\usepackage{enumitem}

\title{Clinical BERTScore: An Improved Measure of Automatic Speech Recognition Performance in Clinical Settings}

\author{Joel Shor \\
    Verily Life Sciences, USA \\
    \texttt{joelshor@verily.com}
    \And
    Ruyue Agnes Bi \\
    MIT, USA \\
    \texttt{ruyuebi@mit.edu} 
    \And
    Subhashini Venugopalan \\
    Google Research, USA
    \AND
    Steven Ibara \\
    Verily Life Sciences, USA
    \And
    Roman Goldenberg\\
    Verily Life Sciences, Israel
    \And
    Ehud Rivlin\\
    Verily Life Sciences, Israel
 }

\begin{document}
\maketitle

\begin{abstract}

Automatic Speech Recognition (ASR) in medical contexts has the potential to save time, cut costs, increase report accuracy, and reduce physician burnout. However, the healthcare industry has been slower to adopt this technology, in part due to the importance of avoiding medically-relevant transcription mistakes. In this work, we present the Clinical BERTScore (CBERTScore), an ASR metric that penalizes clinically-relevant mistakes more than others. We demonstrate that this metric more closely aligns with clinician preferences on medical sentences as compared to other metrics (WER, BLUE, METEOR, etc), sometimes by wide margins. We collect a benchmark of 18 clinician preferences on 149 realistic medical sentences called the Clinician Transcript Preference benchmark (CTP) and make it publicly available\footnote{\url{https://osf.io/tg492/}} for the community to further develop clinically-aware ASR metrics. To our knowledge, this is the first public dataset of its kind. We demonstrate that CBERTScore more closely matches what clinicians prefer.
\end{abstract}

\section{Introduction}

Clinicians in a number of disciplines work in an overburdened healthcare system that leads to difficult working environments and an epidemic of physician burnout \cite{dzau2018care}. AI-related technologies have the potential for improving efficiency on repetitive tasks, therefore increasing both patient throughput and decreasing physician burnout. For example, physicians in a number of disciplines spend as much time doing paperwork as with patients~\cite{tai2017electronic}. However, the adoption of speech technology in the medical community has been slow~\cite{9133298}, and there are a number of speech technologies that could improve efficiency.

Speech technology can be applied to a number of medical problems including transcribing patient-physician conversations~\cite{medical_scribe}, helping dysarthric patients communicate~\cite{shor20_interspeech}, and diagnosing medical conditions from speech~\cite{cap12, Shor_2022, frill, venugopalan21_interspeech}. In this work, we focus on the task of generating a report after a colonoscopy procedure.

One of many reasons for the lower adoption of time-saving speech transcription technologies is that the ASR systems often don't perform as well in real-world clinical settings as they do on evaluation benchmarks. The most common metric for measuring ASR performance, Word Error Rate (WER), has significant practical drawbacks \cite{2003WangWER, morris2004and, 2011HeWER}. First, all mistakes are treated equally. In clinical settings, however, medical words are more important (e.g. "had complete resection" $\rightarrow$ "had complete \textbf{c-section}" is a worse mistake than $\rightarrow$ "\textbf{has} complete resection", but both have equal WER). Second, some mistakes affect the overall intelligibility more than others (e.g. "was no perforation" $\rightarrow$ "was no \textbf{puffer age}" vs "was \textbf{not any} perforation"). Although researchers have proposed alternatives to the WER, no metric combines medical domain knowledge with recent AI advances in language understanding.

In this work, we make the following contributions:

\begin{enumerate}[noitemsep,leftmargin=*]

\item Generate a collection of realistic medical sentences and transcripts with plausible ASR errors and collect preferences from 18 clinicians on 149 sentences. We publicly released this dataset for reference and future studies. This is the first public dataset of its kind.

\item Present the Clinical BERTScore (CBERTScore) and demonstrate that it more closely matches clinician preferences on medical transcripts than other ASR metrics (WER, BLEU, METEOR, BERTScore).

\item Demonstrate that CBERTScore does not perform worse than other metrics on non-medical transcripts.

 \end{enumerate}

\section{Related work}

\begin{figure*}[t]
    \centering
  
  \includegraphics[width=0.3\textwidth]{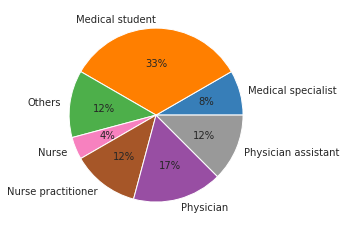}
  \includegraphics[width=0.58\textwidth]{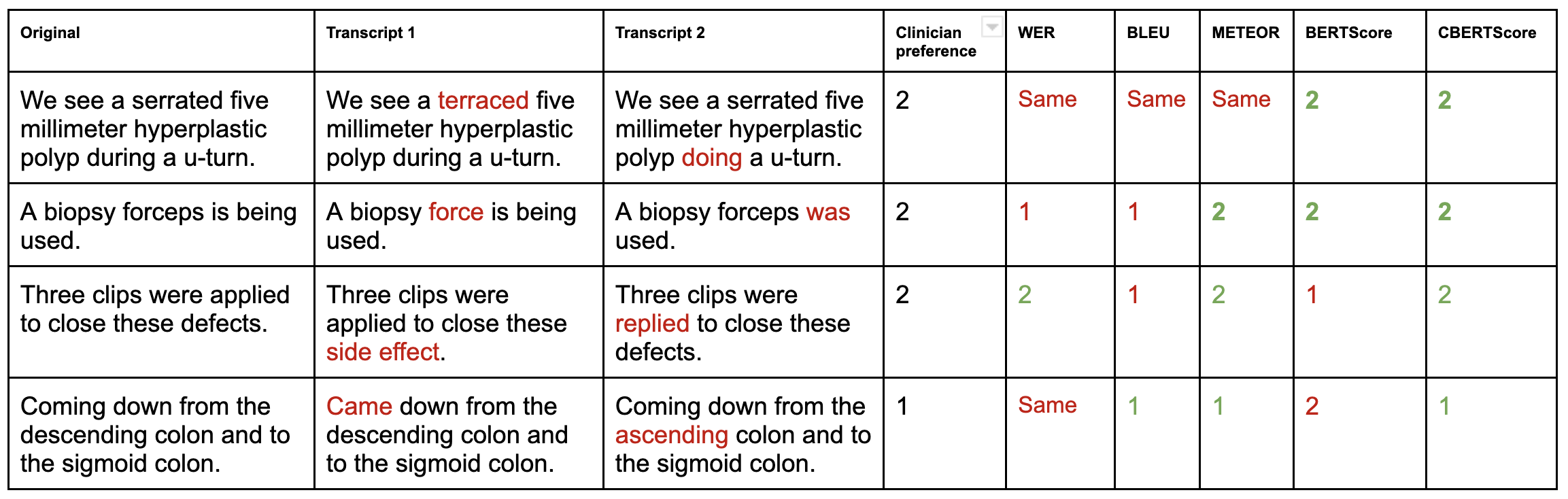}
  
  \caption{\footnotesize{\textbf{Left:} Background of the clinicians who were surveyed to create the Clinician Transcript Preference (CTP) dataset. \textbf{Right:} Some examples of triplet medical sentences, which transcript clinicians prefer, and which transcript scores better based on different metrics.}
  \label{fig:intro}
  }
\end{figure*}

There are a number of ways to evaluate transcript quality. The Word Error Rate (WER), is the simplest to compute and most common. It counts the number of insertions, deletions, and substitutions between two text strings, and normalizes by the length of the reference string. The Bilingual Evaluation Understudy (BLEU) \cite{papineni2002bleu} measures the amount of n-gram overlap between two text strings (where \textit{n} is often 4). It captures the intuition that groups of words are important in addition to individual words. METEOR \cite{banerjee2005meteor} focuses on unigrams, but computes an explicit alignment between two strings and takes both precision and recall into consideration. While these techniques are cheap to compute, they primarily focus on character or string similarity, not semantic similarity.

Our work most closely follows the BERTScore \cite{2019ZhangBERTScore}. This metric computes a neural word embedding for each word in the reference and candidate. Embeddings are matched using cosine distance instead of string similarity, and the final score takes precision and recall into account (see Fig.\ref{fig:intro}). This method takes semantic similarity into account, but not that some words are more important to preserve in clinical contexts.

Structured graphs are one way to encode real-world knowledge in a machine-readable format. The Knowledge Graph (KG) \cite{kg_blog} is a publicly available structure that encodes medical knowledge. Previous work has used the medical subset of the KG to learn medical entity extraction \cite{medical_scribe}. We primarily follow this approach to determine which words are clinically significant.

\section{Methods}

\subsection{Clinical BERTScore}
\label{sec:cbert_def}

Our proposed metric, the Clinical BERTScore (CBERTScore), combines the BERTScore \cite{2019ZhangBERTScore} and the medical subset of the Knowledge Graph \cite{medical_scribe}. 

BERTScore is a relatively novel language generation evaluation metric proposed in \cite{2019ZhangBERTScore} based on pre-trained BERT contextual embeddings. It is designed to capture semantic similarity between two sentences, instead of simple string matching. Given a reference sentence $x = \langle x_1, ..., x_k\rangle$ and a candidate sentence $\hat{x} = \langle \hat{x}_1, ..., \hat{x}_l\rangle$, we first represent each token by a contextual embedding, and then calculate the cosine similarities between the tokens. Each token in the reference sentence is matched to the most similar token in the candidate sentence, and vice versa. The former is used to compute the recall $R_{\textsc{bert}}$, and the latter to compute the precision $P_{\textsc{bert}}$. Precision and recall are then combined into a single score BERTScore as follows:

{\small
\begin{align*}
R_{\textsc{bert}} & = \frac{1}{|x|} \sum_{x_i \in x} \max_{\hat{x}_j \in \hat{x}} \mathbf{x}_i^T \hat{\mathbf{x}}_j , \\
 P_{\textsc{bert}} & = \frac{1}{|\hat{x}|}\sum_{\hat{x}_j \in \hat{x}} \max_{x_i \in x} \mathbf{x}_i^T\hat{\mathbf{x}}_j \\
\text{BERTScore} & = 2\frac{P_{\textsc{bert}} \cdot R_{\textsc{bert}}}{P_{\textsc{bert}} + R_{\textsc{bert}}}
\end{align*}

Building on this insight, we define CBERTScore as:

{\small
\begin{align*}
\textbf{CBERTScore}(x, \hat{x}) = 
& k \times \text{BERTScore}_{\text{medical}}(x, \hat{x}) + \\
& (1-k) \times \text{BERTScore}_{\text{all}}(x, \hat{x}) \\ & \text{, where $0 \leq k \leq 1$}
\end{align*}%
}

BERTScore\textsubscript{all} is computed over all words in the sentences, and BERTScore\textsubscript{medical} is computed over a subset of them that are medically relevant. If there are no medical terms in either the reference or candidate sentence, we define the CBERTScore to be the standard BERTScore (on all words), i.e., $k$ is set at $0$.

We inject medical information into this metric in two ways. First, we compute a weighted score on a subset of words involving medical terms, as determined by the Knowledge Graph \cite{medical_scribe}. Second, we tune the weight of the clinical term penalty to best match a clinician transcript dataset (CTP) that we collected. We describe our method for determining $k$ in Sec. \ref{sec:tuning}.

\subsubsection{Medical Entities}
\label{subsubsec:medical_entities}
Similar to \cite{medical_scribe}, we derive roughly 20K medically relevant words from Google's Knowledge graph \cite{kg_blog}. These words come from entities with properties such as ``/medicine/disease", ``/medicine/drug", ``/medicine/medical\textunderscore treatment", and ``/medicine/medical\textunderscore finding". We also include numbers for the CBERTScore algorithm, since numerical accuracy is important in medical contexts.

\subsubsection{Tuning the medical entities weight factor}
\label{sec:tuning}
CBERTScore has a parameter controlling the weight of the clinical component. To determine this factor, we picked the best performing $k$ on the training subset of the Clinician Transcript Preference (CTP) dataset (Sec. \ref{subsubsec:ctp}). We evaluated $k$ using 11 points evenly spaced between 0 and 1, and performed the evaluation methodology in Sec. \ref{subsubsec:ctp} for each. We then used this value for all subsequent results and analyses.

\subsection{Clinician Transcript Preference (CTP) Dataset}
\label{subsubsec:ctp}
In order to compare CBERTScore's agreement with human preference, we sent out a Qualtrics survey to elicit judgment specifically from clinicians\footnote{Broadly defined as a person with extensive clinical experience or from a clinical research background, for our purpose.}. We call this dataset the Clinician Transcript Preference dataset (CTP), and we make it publicly available on the Open Science Framework (OSF). To our best knowledge, this is the first publicly available dataset with clinician preferences of transcript errors.

We collected data on 150 sentences. They were divided into three groups, each containing 50 trials. 18 subjects with clinical backgrounds responded to more than half the questions. Fig. \ref{fig:intro} (left) describes clinician backgrounds. Each participant was randomly assigned to a group to ensure approximately uniform response coverage. For each trial, participants are given a ground truth sentence and two ``transcripts" and asked to select the less useful one or to indicate the two are about the same. An example of such a triplet is as follows:
\textit{``Patient elects to go under Propofol sedation."}
\begin{tabular}{p{0.5cm}l}
    \textbf{\#1}:&Patient elects to go under Prilosec sedation. \\
    \textbf{\#2}:&Patient selects to go under Propofol sedation. 
\end{tabular}
The survey was designed to take no more than 20 min to minimize the cognitive strain on participants. One sentence was malformed, resulting in 149 sentences for the final dataset.


\begin{figure*}[t]
    \centering
  \includegraphics[width=0.8\textwidth]{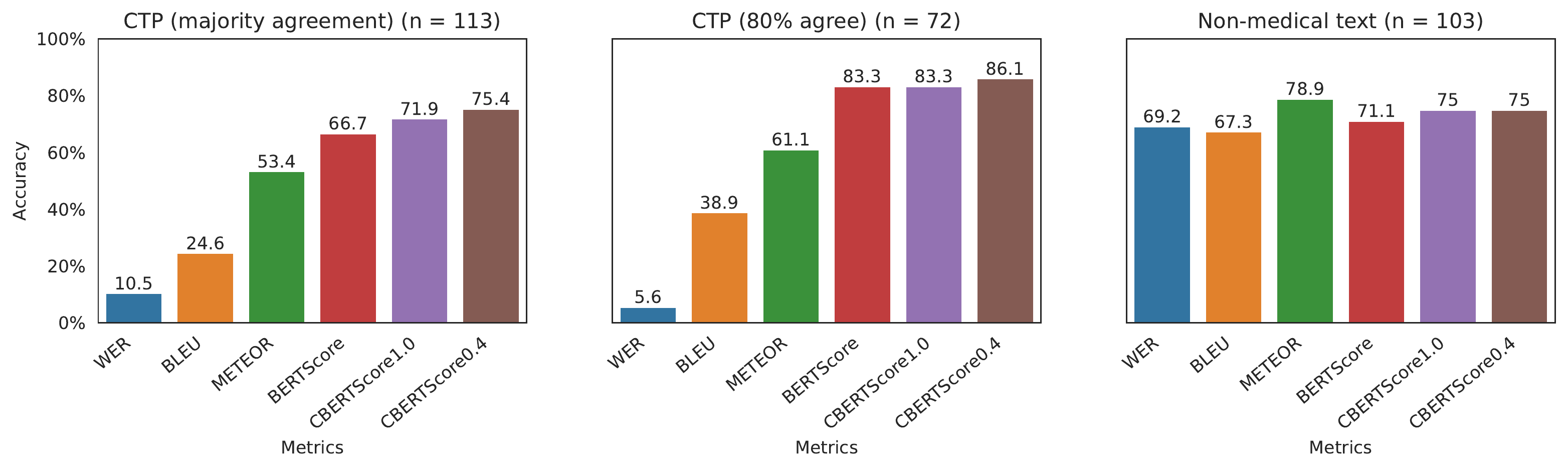}
  
  \caption{\footnotesize{Comparison of different metrics' agreement with human rater transcript preferences. Process of deriving a prediction from metric values is described in Sec. \ref{subsubsec:ctp}. In all plots, "CBERTScore1.0`` is the performance from only the medical term component ($k=1.0$ in Sec. \ref{sec:cbert_def}). "CBERTScore0.4`` uses the optimal value of $k$ according to the train set. \textbf{Left:} Agreements with clinicians on the CTP benchmark when labels are derived using majority voting. \textbf{Center:} Agreements with clinicians on the CTP benchmark when restricted to questions with unanimous answers. \textbf{Right:} Agreement with speech pathologist raters on the non-medical dataset, when restricting the data to cases where there is a fidelity difference between two candidate transcripts.}}
  \label{fig:results}
\end{figure*}

\subsubsection{Constructing the CTP triplets}

To generate the triplets of (target, transcript \#1, transcript \#2) used in the survey, we started by downloading publicly available YouTube videos on colonoscopies created by GI physicians and educational institutes. The target sentences were transcribed by Google's publicly available Speech-to-Text medical dictation model \cite{Soltau2021UnderstandingMC} and manually checked for accuracy. Filler words such as ``uh" and repeated words were edited out. Sentences longer than 30 words or less than 5 were discarded. 

For each target sentence, transcript \#1 was generated by one of Google's other, non-medical, publicly available ASR models. Transcripts with an edit distance\cite{edit_distance} outside [1, 3] were discarded. This procedure generated 1220 candidate sentences. 

To ensure that the two transcripts were roughly comparable in terms of fidelity, transcript \#2 was generated synthetically. We used a publicly available English word frequency dictionary\cite{freq_dict} to select words in the target sentence that were candidates for synthetic errors. Candidate words were at least 5 characters, appeared in the 1M word dictionary fewer than 10 times, and were not proper nouns. 486 candidate sentences matched these criteria. Finally, transcript \#2 was generated by deleting the candidate word or manually substituting it with a phonetically similar word or phrase\footnote{A Python fuzz search algorithm based on CMU Pronouncing Dictionary was used for consistency.}. We discarded similar sentences and selected 150 triplets for the final survey. The ordering of the two transcripts was randomized, and so were the sentences.

\subsubsection{Evaluating metrics on the CTP}
\label{sec:eval_method}

To compare the ability of different metrics to agree with rater preference from the CTP, we define a 3-class classification problem as follows:
\begin{align*}
\text{Predicted}&\text{ better transcript}(M)(gt, t_1, t_2) = \\
&\begin{cases} 
t_1         & \, M(gt, s_1) - M(gt, s_2) > l \\
t_2         & \, M(gt, s_1) - M(gt, s_2) < -l \\
\text{same} & \, \text{else}
\end{cases} 
\end{align*}
where $M$ is an evaluation metric, $gt$ is the ground truth sentence, and $t_i$ are the transcripts. Note the predictions are reversed for the WER, since lower values indicate higher fidelity. $l$ is a free variable, which we optimize separately for each metric. We split the data into two halves, choose the best performing $l$ on one half, and report the accuracy using that $l$ on the second half.

\subsubsection{Non-medical sentences}
To demonstrate that CBERTScore doesn't degrade on non-medical speech, we compare the metrics' agreement with rater preferences on a dataset with annotations similar to \cite{tobin22_s4sg}. Part of this dataset consists of 5-tuples of (ground truth sentence, transcript 1, transcript 2, assessment 1, assessment 2), where the sentence assessments describe how much of the ground truth sentence's meaning is captured in the transcript. We used a subset of 103 utterances from our annotated data where the ratings were not the same, and at least one transcript was rated as having ``Major errors". We report performance using a similar formulation as on the CTP evaluation in Sec. \ref{sec:eval_method}: we frame this as a 2-way classification problem (no cutoff is needed since we exclude tuples that have the same rating).

\section{Results}

\subsection{Clinician responses}

18 clinicians responded to a total number of 149 triplet questions. Each question had 5 or 6 responses. 78\% of questions had more than half agreement on which transcript was less useful and 42\% had more than 80\% agreement. Clinicians thought transcripts were the same usefulness in 21\% of cases.

\subsection{Metric agreement on medical text}

We report 3-way accuracy classification on the CTP dataset using two labeling schemes (Fig. \ref{fig:results}). In the first, we only look at the questions where more than half the respondents agreed. In the second, we report accuracy on the questions where more than 4/5 of the respondents agreed. For both numbers, we determine the cutoff from one half the data and report accuracy on the second half. 

First, the metric ordering by performance is the same using both labeling schemes, and the best CBERTScore medical weighting factor was the same using both label schemes. Second, BERTScore and CBERTScore are significantly more closely aligned with clinician preferences than other metrics. Third, CBERTScore weighted entirely toward medical terms outperforms or ties with BERTScore agreement. Fourth, the weighted combination of medical and non-medical terms outperforms other metrics in terms of clinician agreement. Fifth, the medical component meaningfully improves the performance of CBERTScore over BERTScore (75.9\% vs 67.2\% and 87.5\% vs 84.4\%).

\subsection{Metric agreement on non-medical text}

CBERTScore was the second best-performing metric on non-medical text. Importantly, the addition of the medical component did not degrade the performance compared to BERTScore.

\section{Discussion}

\subsection{Knowledge Graph medical terms wins and losses on the CTP}

The CTP (Sec. \ref{subsubsec:ctp}) had 127 distinct words that were the source of transcript errors, and 684 distinct other words. The medically-relevant terms used in the CBERTScore algorithm, identified primarily from the Knowledge Graph as described in Sec. \ref{subsubsec:medical_entities}, intersected with 99 of the 127 transcript error words. By manual inspection, 25 of the 28 transcript error words in the CTP not included in the CBERTScore word list were used in a medical context but were not only medical in meaning (ex. ``surveillance", ``tethered", and ``longitudinal"). 3 of the 28 missed words did have a primarily medical meaning, but were not included in the CBERTScore list either due to errors in the KG or errors in the queries generating the list (``cologuard", ``colonoscope", ``protuberance"). Some of the words have a clear meaning in a medical context, and could be manually added to the list for future applications (``snare", ``suctioning", etc.).

The CBERTScore word list included 100 words that weren't selected for transcript errors. Many of these are medical in nature, but were not selected for synthetic transcript errors via the method described in Sec. \ref{subsubsec:ctp} (ex. ``endoscope", ``hypoplastic", ``lymphoma").

\subsection{CBERTScore performance on the CTP}

\begin{figure}[t]
    \centering
    \includegraphics[width=0.4\textwidth]{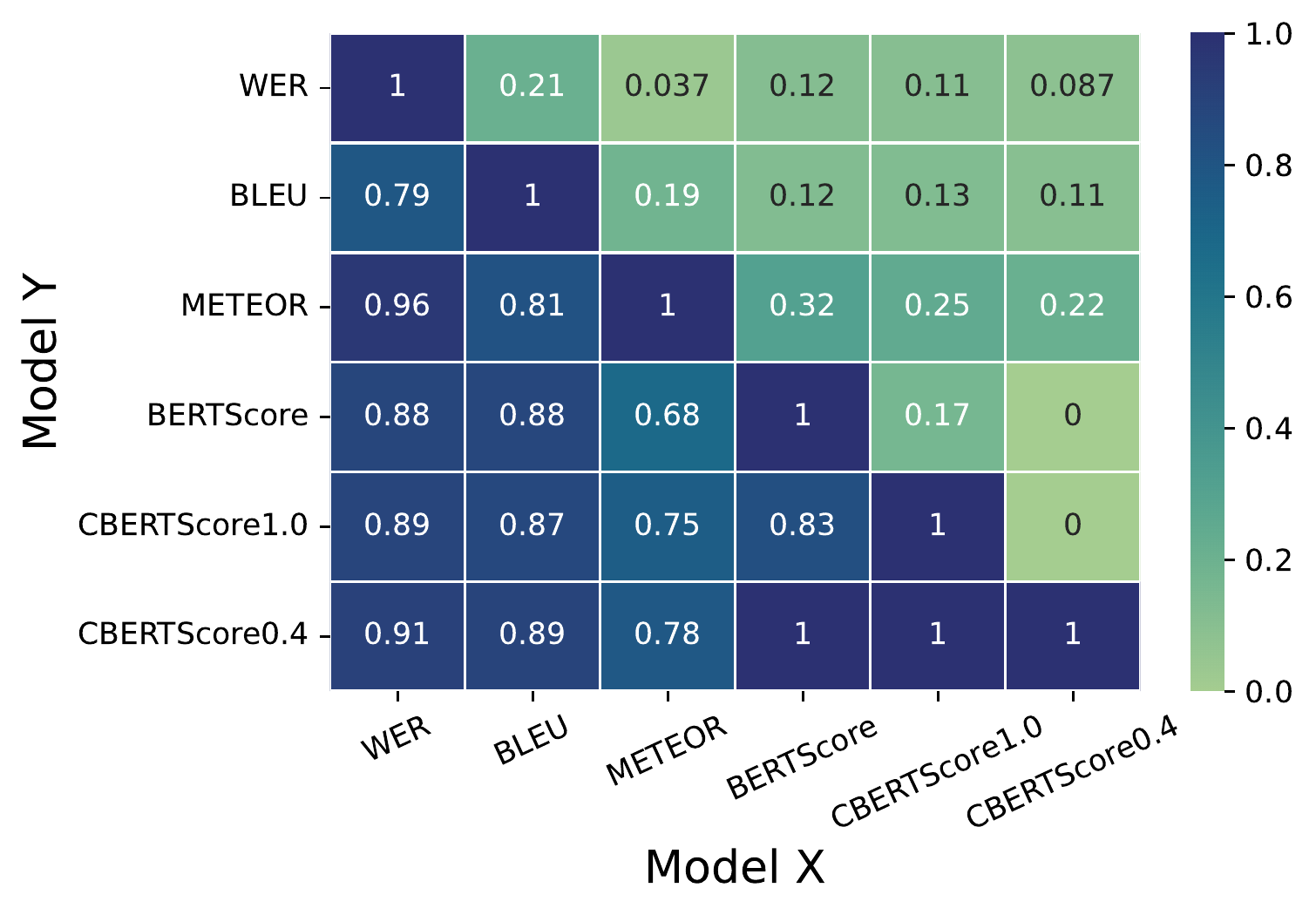}
    \captionof{figure}{\footnotesize{Fraction of cases where metric Y is correctly conditioned on metric X and Y disagreeing. An indicator of how similar the pattern of mistakes is between metrics.}}
  \label{fig:discussion}
\end{figure}

\subsubsection{CBERTScore wins}

Fig. \ref{fig:discussion} left shows the degree to which better-performing metrics subsume other metrics, or make a different pattern of mistakes. The plot shows the $(\text{Metric Y correct})/(\text{Metric X and Y disagree)}$. Metrics that have higher clinician agreement and a high fraction on this plot are strictly better, whereas metrics with higher agreement but a low value in this plot indicate that another metric might have an additional signal. We see that CBERTScore is nearly strictly better than the other metrics, with the possible exception of METEOR (when they differ, METEOR gives the correct rating in roughly a third of cases).

There were some triplets that CBERTScore got correct that no other metric did. The improvements over BERTScore always involved a medical term, and sometimes involved encouraging the metric to prioritize medical mistakes (ex. "Marked the site with 5 cc\textquotesingle s of indigo carmine.`` $\rightarrow$ "Marked the site with 5 \textcolor{red}{cici\textquotesingle s} of indigo carmine.`` vs "Marked the \textcolor{red}{sight} with 5 cc\textquotesingle s of indigo carmine.``)

There were thirteen triplets that the neural word embeddings predicted correctly that other metrics did not. Many of these wins came from the strength of neural word embeddings penalizing less for semantically similar mistakes (ex. "Small burst of coagulation to create a darkish white ablation.`` $\rightarrow$ "Small burst of coagulation to create a darkish white \textcolor{red}{oblation}.`` vs "Small burst of coagulation to create a \textcolor{red}{dark} white ablation.``). Furthermore, BERTScore agreed with clinicians on some medical word mistakes, likely due to the BERT embedding somewhat understanding when a transcript error leads to a large semantic change in a medical term (ex. "No ongoing infection or coagulopathy.`` $\rightarrow$ "No \textcolor{red}{on going} infection or coagulopathy.`` vs "No ongoing infection or \textcolor{red}{glomerulopathy}.``).

\subsubsection{CBERTScore mistakes}

Fig. \ref{fig:discussion} shows that METEOR made the most correct predictions when CBERTScore was incorrect. Some mistakes are due to the KG medical list being incomplete. For example, "longitudinal`` was not included, but has medical meaning in clinical contexts (ex. "The longitudinal extent of the hot snare.`` $\rightarrow$ "The \textcolor{red}{long eternal} extent of the hot snare.`` vs "The longitudinal \textcolor{red}{extend} to the hot snare.``).

Another pattern of mistake is when a non-medical adjective contains an error, but the adjective modifies a medical term in an important way. For example, "vessel`` is a medical term, but "feeding`` is not (ex. "This polyp is at high risk of bleeding, with multiple feeding vessels.`` $\rightarrow$ "This polyp is at high risk of bleeding, with multiple \textcolor{red}{seeding} vessels.`` vs "This polyp is at high \textcolor{red}{risking} bleeding, with multiple feeding vessels.``). This suggests that future work might include modifications and dependencies when calculating clinical importance.

Finally, a third pattern of mistake involves the fact that METEOR penalizes complex correspondences between candidate and reference sentences, while CBERTScore only considers the best pairwise word matches. One example in the CTP preserves most of the words, but reorders them (ex. "Inject into the head of the polyp, another 1 to 2 cc.`` $\rightarrow$ "\textcolor{red}{Injectant} the head of the polyp, another 1 to 2 cc.`` vs "Inject into the head of the polyp, another 1 \textcolor{red}{2 to} cc.``).

\section{Conclusions}

We present CBERTScore, a novel metric that combines medical domain knowledge and recent advances in neural word embeddings. We collect and release a benchmark of clinician rater preferences on transcript errors, demonstrate that CBERTScore is more closely aligned with clinician preferences, and release the benchmark for the research community to continue to improve ASR in medical contexts.

\bibliography{anthology,mybib}
\bibliographystyle{acl_natbib}

\end{document}